\def\papertitle{Leveraging Real Electric Guitar Tones and Effects to Improve Robustness in Guitar Tablature Transcription Modeling}
\def\paperauthorA{Hegel Pedroza}
\def\paperauthorB{Wallace Abreu}
\def\paperauthorC{Ryan M. Corey}
\def\paperauthorD{Iran R. Roman}

\PassOptionsToPackage{pdftex}{graphicx}
\documentclass[twoside,a4paper]{article}
\usepackage{etoolbox}
\usepackage{dafx_24_lbr}
\usepackage{amsmath,amssymb,amsfonts,amsthm,enumitem}
\usepackage{euscript}
\usepackage[T1]{fontenc}
\usepackage[utf8]{inputenc}
\usepackage{ifpdf}
\usepackage[english]{babel}
\usepackage{caption}
\usepackage{subfig} 
\usepackage{color}
\usepackage{tabularx}
\usepackage{cite}

\usepackage{graphicx}
\usepackage{booktabs}

\input glyphtounicode
\ninept

\newcounter{numauth}
\newcounter{listcnt}
\newcommand\authcnt[1]{\ifdefined#1 \stepcounter{numauth} \fi}

\newcommand\addauth[1]{
\ifdefined#1 
\stepcounter{listcnt}
\ifnum \value{listcnt}<\value{numauth}
\appto\authorslist{, #1}
\else
\appto\authorslist{~and~#1}
\fi
\fi}
\authcnt{\paperauthorB}
\authcnt{\paperauthorC}
\authcnt{\paperauthorD}
\authcnt{\paperauthorE}
\authcnt{\paperauthorF}
\authcnt{\paperauthorG}
\authcnt{\paperauthorH}
\authcnt{\paperauthorI}
\authcnt{\paperauthorJ}
\def\authorslist{\paperauthorA}
\addauth{\paperauthorB}
\addauth{\paperauthorC}
\addauth{\paperauthorD}
\addauth{\paperauthorE}
\addauth{\paperauthorF}
\addauth{\paperauthorG}
\addauth{\paperauthorH}
\addauth{\paperauthorI}
\addauth{\paperauthorJ}

\usepackage{times}

\newif\ifpdf
\ifx\pdfoutput\relax
\else
   \ifcase\pdfoutput
      \pdffalse
   \else
      \pdftrue
   \fi
\fi

\ifpdf 
  \usepackage[pdftex,
    pdftitle={\papertitle},
    pdfauthor={\authorslist},
    pdfsubject={Proceedings of the 27th International Conference on Digital Audio Effects (DAFx24)},
    colorlinks=false, 
    bookmarksnumbered, 
    pdfstartview=XYZ 
  ]{hyperref}
  \usepackage[pdftex]{graphicx}
\else 
  \usepackage[dvips]{epsfig,graphicx}
  \usepackage[dvips,
    pdftitle={\papertitle},
    pdfauthor={\authorslist},
    pdfsubject={Proceedings of the 27th International Conference on Digital Audio Effects (DAFx24)},
    colorlinks=false, 
    bookmarksnumbered, 
    pdfstartview=XYZ 
  ]{hyperref}
\fi
\usepackage[hypcap=true]{caption}
\title{\papertitle}

\fouraffiliations{
\paperauthorA}
{
National Autonomous University of Mexico\\ Mexico City, Mexico
}
{\paperauthorB}
{
Federal University of Rio de Janeiro \\ Rio de Janeiro, Rio de Janeiro, Brazil 
}
{\paperauthorC}
{
Discovery Partners Institute \& University of Illinois Chicago \\ Chicago, Illinois, USA 
}
{\paperauthorD}
{
New York University \\ New York, New York, USA 
}

\begin{document}
\ifpdf 
  \DeclareGraphicsExtensions{.png,.jpg,.pdf}
\else  
  \DeclareGraphicsExtensions{.eps}
\fi


\maketitle

\begin{abstract}

\noindent Guitar tablature transcription (GTT) aims at automatically generating symbolic representations from real solo guitar performances. 
Due to its applications in education and musicology, GTT has gained traction in recent years.
However, GTT robustness has been limited due to the small size of available datasets.
Researchers have recently used synthetic data that simulates guitar performances using pre-recorded or computer-generated tones, allowing for scalable and automatic data generation.
The present study complements these efforts by demonstrating that GTT robustness can be improved by including synthetic training data created using recordings of real guitar tones played with different audio effects. 
We evaluate our approach on a new evaluation dataset with professional solo guitar performances that we composed and collected, featuring a wide array of tones, chords, and scales.


\end{abstract}

\section{Introduction}
\label{sec:intro}

Guitar tablature transcription (GTT), a form of automatic music transcription (AMT)~\cite{benetos2018automatic,Gowrishankar2016amtreview}, involves transcribing real guitar performances into tablatures~\cite{burlet2013robotaba, Peleaari2008Multimodal}.
Unlike standard Western notation, tablatures intuitively illustrate finger placements, and are thus of high relevance for music education~\cite{thompson2011speaking,harrison2010challenges}, musicological research~\cite{gavito2015oral}, guitar performance theory~\cite{de2022guitar, de2018fretboard, koozin2011pop}, and general communication of artistic expression.

Significant advances in GTT have been achieved through deep learning models, primarily trained and evaluated using the GuitarSet dataset~\cite{xi2018guitarset, 
wiggins2019guitar, Cwitkowitz2023Fretnet, kim2022attentionMechanism, bastas2022inharmonicity, byambatsogt2020robots, riley2024high, Jadhav2022}. However, its limited size has led to poor generalization capabilities~\cite{Zang2024synthtab}. 
This is the well-known ``domain-shift problem'' in machine learning, which explains model failures in real-world applications due to discrepancies between the training conditions and actual usage environments~\cite{quinonero2022dataset}.
Therefore, assessment of GTT model robustness in new domains is essential to understand their usefulness.
Data augmentation is a common technique to mitigate ``domain-shift''. 
By systematically modifying existing data or metadata, additional examples can be generated and used as training data, thereby expanding a model’s familiarity with a wider data distribution~\cite{shorten2019survey}. 
This technique has recently shown improvements in related tasks like sound event detection~\cite{salamon2017scaper,roman2024spatial,roman2024enhanced,dinkel2021towards}.

In the case of GTT, data augmentation can be carried out by taking existing tablatures to guide the temporal placement of pre-recorded or synthesized individual guitar tones to simulate a guitar performance. Zang \textit{et al.}~\cite{Zang2024synthtab} recently used this technique with commercially-available guitar synthesizers. 


We want to investigate whether diversifying the timbres present in training data by including audio effects improves GTT. Since recording and annotating solo guitar with audio effects is highly time-intensive, we draw inspiration from Zang et al.’s SynthTab~\cite{Zang2024synthtab} to scalably simulate guitar performances from guitar tablature. However, instead of using synthetic guitar tones (i.e. ``MIDI''), we hypothesize that GTT robustness can be enhanced by two factors: the exclusive use of real guitar tones and audio effects to generate synthetic data for model training.

Finally, after training we assess model robustness using a new dataset of professional solo guitar performances that we composed and collected for this study.
This new dataset features a diverse array of performance styles. Therefore,
our key contributions are:
\begin{enumerate}[nolistsep]
\item A scalable method to generate data to train GTT models using real recordings of guitar tones and audio effects.
\item A new dataset for evaluating GTT models.
\item A benchmark clearly demonstrating the benefits of our approach for model  robustness.
\end{enumerate}
We release our code to reproduce our data generation and model training, as well as the new dataset we collected for this study\footnote{\href{https://robust-guitar-tabs.github.io}{\texttt{robust-guitar-tabs.github.io.}}}.

\begin{table*}[ht]
\centering
\caption{TabCNN performance on GuitarSet. Averages ($\pm$ denotes the standard deviation) across the six hold-out folds. Top row: metrics by Wiggins \& Kim~\cite{wiggins2019guitar}  (we could reproduce their results). Bottom rows: outcomes when training data includes simulated tracks.}

\label{tab:metrics}
\begin{tabularx}{\textwidth}{@{}l *{7}{>{\centering\arraybackslash}X}@{}}
\toprule
 & \multicolumn{3}{c}{\textbf{Multi-pitch estimation}} & \multicolumn{3}{c} {\textbf{Tablature estimation}} & \\
\cmidrule(lr){2-4} \cmidrule(lr){5-7}
& \textbf{F\textsubscript{1}} & \textbf{P} & \textbf{R} & \textbf{F\textsubscript{1}} & \textbf{P} & \textbf{R} & \textbf{TDR} \\ \midrule
TabCNN\cite{wiggins2019guitar} & 0.826$\pm$0.025 & 0.900$\pm$0.016 & 0.764$\pm$0.043 & \textbf{0.748}$\pm$0.047 & \textbf{0.809}$\pm$0.029 & 0.696$\pm$0.061 & 0.899$\pm$0.033 \\
\hline
+ GuitarSetFX & \textbf{0.837}$\pm$0.019 & 0.904$\pm$0.019 & \textbf{0.785}$\pm$0.038 & 0.746$\pm$0.030 & 0.795$\pm$0.022 & \textbf{0.708}$\pm$0.044 & 0.896$\pm$0.021 \\
+ GuitarProFX & 0.830$\pm$0.018 & \textbf{0.908}$\pm$0.014 & 0.769$\pm$0.027 & 0.743$\pm$0.029 & 0.802$\pm$0.029 & 0.696$\pm$0.035 & \textbf{0.900}$\pm$0.021 \\
\bottomrule
\end{tabularx}
\end{table*}

\section{Methods}

\subsection{Datasets for training and validation}

Like Wiggins \& Kim~\cite{wiggins2019guitar}, we train TabCNN using GuitarSet~\cite{xi2018guitarset} for training and cross-validation. Additionally, we introduced two synthetic datasets to expand the data used for model training: GuitarSetFX and GuitarProFX.
Our approach maximizes tone diversity by randomly selecting tones from a large set of pre-recorded single guitar notes.
Thus, in our synthetic solo guitar performances, melodies and chords consists of notes played with varied guitar tones and audio effects. 
This data generation strategy aligns with our hypothesis that such diversity will enhance the model’s robustness, allowing it to concentrate on pitch content and guitar string+fret inference, while disregarding specific timbre qualities.

The guitar tone set we use for GuitarSetFX and GuitarProFX includes the clean tones and effects from EGFxSet\footnote{We exclude the tones processed through the ``delay'' effect due to audibly repeated tone onsets that would confuse our model.}, which used a 2004 Fender Stratocaster guitar~\cite{pedroza2022egfxset}. 
Additionally, we recorded all possible clean notes in a 1978 Ibanez Performer 300 guitar, using alternate-picking technique and captured through direct input via an Audient iD14 audio interface, also miked through an Orange CR-60 amplifier with a Shure sm57 microphone.
The amplifier settings were adjusted with bass and treble knobs at 5 and plate reverb at 3.

GuitarSetFX reproduces the 360 guitar performance tracks of GuitarSet, while GuitarProFX comprises 360 randomly-chosen solo performance tracks from DadaGP~\cite{sarmento2021dadagp}.
All resulting audio clips are resampled to the 22050Hz sampling rate expected by TabCNN.

\subsection{Model training, and validation}

The GTT model we train is TabCNN, using the implementation in AMT tools~\cite{Cwitkowitz2023Fretnet}\footnote{\href{https://github.com/cwitkowitz/amt-tools}{github.com/cwitkowitz/amt-tools.}}. More specifically, we train three different TabCNN models. The first one is a reproduction of the six-fold cross-validated training setup by Wiggins \& Kim~\cite{wiggins2019guitar}.
The second one follows the same cross-validation but duplicates the training data using the corresponding synthetic GuitarSetFX tracks. The third one also follows the same cross-validation but adds all the synthetic GuitarProFX tracks to the training split.
All other aspects of the training setup remain the same as in the original TabCNN implementation, including model architecture, optimizer, learning rate, batch size, and validation data~\cite{wiggins2019guitar}.

\subsection{The EGSet12 evaluation set}

Furthermore, we introduce EGSet12, a new evaluation set with twelve original solo electric guitar performances (31.65 seconds avg. duration, totaling 379.8 seconds). These pieces were composed by a professional musician and guitar player for this project, showcasing the full tonal range of the electric guitar across diverse melodies and chord complexities. EGSet12 encompasses a broad spectrum of styles, including pop, funk, jazz and twelve-tone, reflecting varied tonalities, keys, rhythms, and modes.

EGSet12 was performed by a single professional guitarist using a Sire T7 Telecaster guitar and a Yamaha B15 amplifier. 
The performance setup allowed the performer to freely select the guitar's volume and tone knobs, also allowing techniques like alternate picking, hybrid picking, and palm mute. 
We captured the performance using an ECM8000 microphone positioned 15 centimeters from the amplifier and connected to a UMC202 HD audio interface (original sampling rate of 48000Hz; resampled to 22050Hz for model inference). 
This recording setup differs significantly from those used in any of the training and validation datasets that we used, offering a new testing domain.
EGSet12, offers realism, tone diversity, and varied playing styles, making it valuable for assessing GTT model robustness. 

EGSet12 features a realistic noisy recording setup and diverse guitar tones and techniques. 
Other than the amplifier, its content was not processed using other guitar effects. 
Future research can process it through more effects to further study model robustness.

\subsection{Metrics}

Consistent with Wiggins \& Kim~\cite{wiggins2019guitar}, we use two types of metrics: multi-pitch and tablature estimation. Multi-pitch metrics assess model performance at the level of pitch estimation and can be thought of as independent of the guitar hardware.
Tablature estimation metrics assess the model's ability to determine which specific string and fret produced a tone. Both are broken down by F\textsubscript{1} score, precision, and recall. Additionally, the tablature disambiguation rate (TDR) calculates how often a correctly-identified pitch gets assigned to the correct fret and string.

\section{Results}

\begin{table*}[ht]
\centering
\caption{TabCNN performance on EGSet12. Each cell is a metric averaged across the twelve tracks ($\pm$ denotes the standard deviation). Top row: performance as trained by Wiggins \& Kim~\cite{wiggins2019guitar}. Bottom rows: performance when training data includes simulated tracks. ``*'' denotes a statistically-significant difference ($p<0.05$ via t-test) compared to the model by Wiggins \& Kim~\cite{wiggins2019guitar}. ``$^\diamond$'' denotes a marginally-significant difference ($0.1>p>0.05$). The underlying distributions of significant (or marginally-significant) comparisons are normal-shaped based on the D’Agostino and Pearson’s test.}

\label{tab:test}
\begin{tabularx}{\textwidth}{@{}l *{7}{>{\centering\arraybackslash}X}@{}}
\toprule
 & \multicolumn{3}{c}{\textbf{Multi-pitch estimation}} & \multicolumn{3}{c}{\textbf{Tablature estimation}} &\\
\cmidrule(lr){2-4} \cmidrule(lr){5-7}
& \textbf{F\textsubscript{1}} & \textbf{P} & \textbf{R} & \textbf{F\textsubscript{1}} & \textbf{P} & \textbf{R} & \textbf{TDR} \\ \midrule
TabCNN\cite{wiggins2019guitar} & 0.638$\pm$0.060 & 0.819$\pm$0.080 & 0.530$\pm$0.067 & 0.447$\pm$0.071 & 0.565$\pm$0.089 & 0.375$\pm$0.067 & 0.695$\pm$0.075 \\
\hline
+ GuitarSetFX & \textbf{0.740}$\pm$0.055$^*$ & 0.835$\pm$0.085 & \textbf{0.679}$\pm$0.052$^\diamond$ & 0.557$\pm$0.088 & 0.619$\pm$0.100$^*$ & 0.518$\pm$0.084 & 0.755$\pm$0.106 \\
+ GuitarProFX & 0.719$\pm$0.061$^\diamond$ & \textbf{0.839}$\pm$0.082 & 0.647$\pm$0.068 & \textbf{0.585}$\pm$0.084 & \textbf{0.658}$\pm$0.073$^*$ & \textbf{0.541}$\pm$0.087$^\diamond$  & \textbf{0.819}$\pm$0.075$^\diamond$ \\
\bottomrule
\end{tabularx}
\end{table*}

\begin{figure*}[ht]
\center
  \setlength{\intextsep}{0pt}
  \setlength{\abovecaptionskip}{2pt}
  \setlength{\belowcaptionskip}{-9pt}
\includegraphics[width=6.9in]{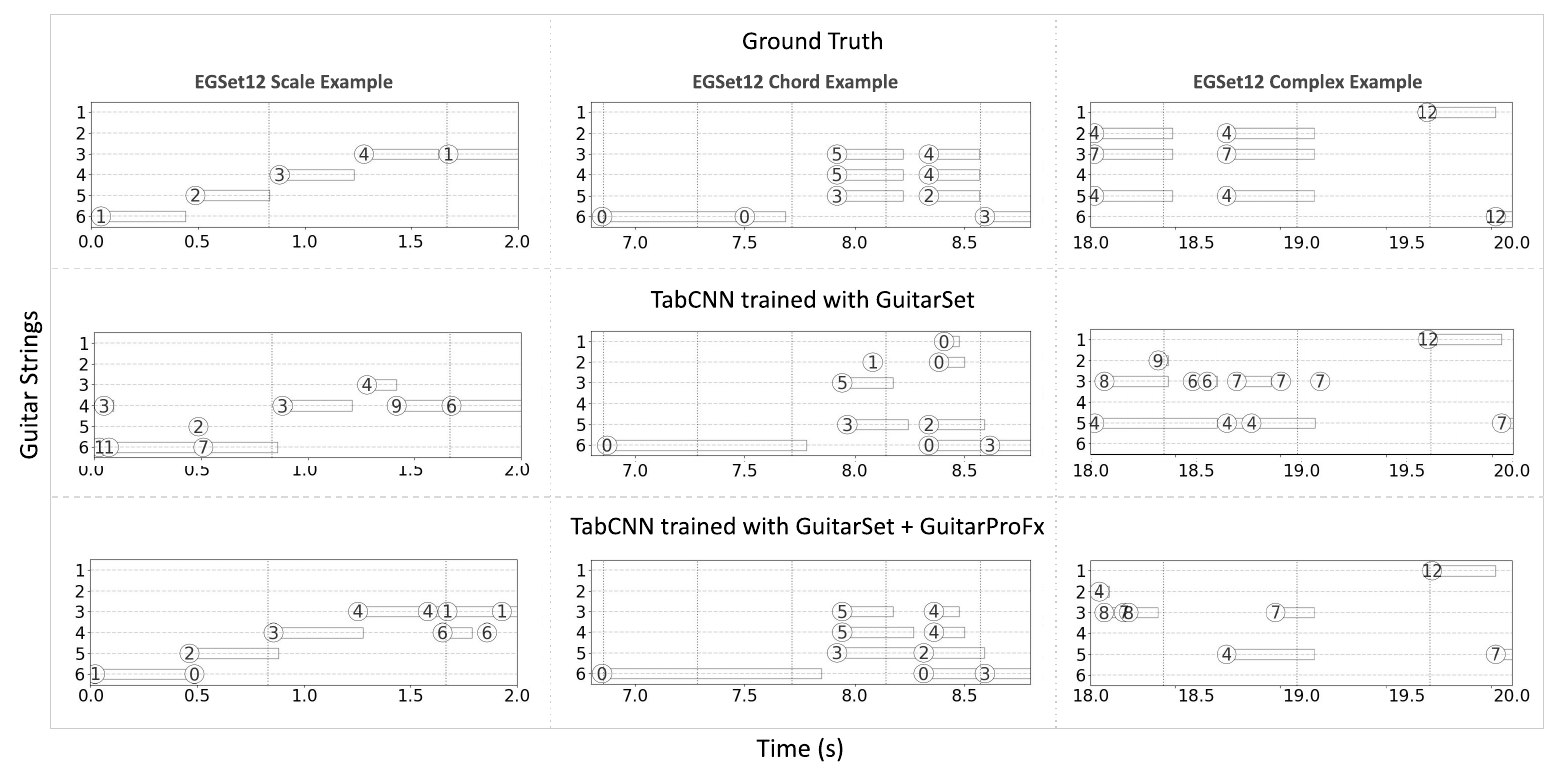}
\caption{\label{detection-examples}{Each column is a two-second EGSet12 excerpt, comparing models trained using GuitarSet, with and without GuitarProFX, against ground truth. Each circled number is a tracked note on a specific guitar fret over time, the vertical lines indicate musical beats.}
}
 
\end{figure*}   

\subsection{GuitarSet cross-validation}

First, we assessed TabCNN performance on GuitarSet during its cross-validated training. 
Table~\ref{tab:metrics} shows the results. 
In general, we observe slight improvements when the training set includes synthetic data with effect tones. 
However, these improvements are minor. 
This indicates that cross-validated performance of TabCNN on GuitarSet does not benefit (or suffer) much from the addition of synthetic data with guitar effects.

\subsection{Model evaluation on the new EGSet12 test domain}

Next, we assessed the three models on EGSet12, with results presented in Table~\ref{tab:test}. 
Although, model performance was comparable during cross-validation with GuitarSet (see Table~\ref{tab:metrics}), evaluation on EGSet12 showed that models trained with synthetic data using audio effects exhibited superior generalization. 
Notably, the 
F\textsubscript{1} score improved more than 10 percentage points for both multi-pitch and tablature estimation, primarily driven by major gains in model recall. 
Precision remained consistent across models for multi-pitch metrics but showed improvements of $\sim$9 points in GuitarProFX tablature estimation and slightly over 5 points in GuitarSetFx. Finally, the TDR also saw improvements of more than 10 points for the model trained using the GuitarProFX data.

\subsection{Qualitative comparison of TabCNN models on EGSet12}

Figure~\ref{detection-examples} shows a qualitative comparison between the TabCNN models we tested on EGSet12. 
The first two columns in Figure~\ref{detection-examples} feature musical excerpts showing that TabCNN trained with GuitarProFx demonstrated greater stability, meaning that it did not exhibit constant string jumping when trying to locate the guitar notes in the audio signal.
Therefore, its predictions of chords and individual notes were more aligned with ground truth. 

The third column is a more challenging example. It shows that 
both models' performance decreased
when processing complex musical passages with harmonic content across multiple simultaneous strings and dissonances such as minor seconds.
Additionally, 
both models struggled to make 
good predictions for the higher frets, as most training data features fret numbers below 12. 
However, in both cases, some pitch predictions were accurate as tablature was associated with possible fingerings in other frets.


\section{Discussion}
Our results in Table \ref{tab:test}, showing the benefits of synthetic training data, are consistent with empirical evidence from the sound event detection literature~\cite{salamon2017scaper,roman2024spatial,roman2024enhanced,dinkel2021towards}.
An interesting observation is the fact that the benefit size was not evident during the cross-validated training (see Table~\ref{tab:metrics}).
This highlights the importance of evaluating GTT on new, structured domains and controlled scenarios (like EGSet12) to assess model robustness accurately.

Another point of discussion is how valid EGSet12 is as a test set for GTT. For example, considering EGSet12 uses an electric guitar, is it fair to use it to evaluate GTT models trained with an acoustic guitar dataset (i.e GuitarSet)?.
We believe that this is the case since in our setup the electric guitar used in EGSet12 and all recording hardware (including microphone, amplifier, and interface) is different from not just GuitarSet, but also GuitarSetFX, and GuitarProFX. 
The fact that tablature estimation precision is higher on EGSet12 for models trained using GuitarSetFX or GuitarProFX (see table~\ref{tab:test}) may suggest an advantage due to the fact that these datasets used electric guitars.
However, the more than 10-point increase in multi-pitch estimation F\textsubscript{1} indicates that model robustness is driven by correct inference of pitch content. 
Therefore, our evidence suggests that EGSet12 is a fair benchmark.

We studied how using real guitar tones processed through audio effects hardware to generate training data improves GTT model robustness. 
Therefore, our current study is inspired by SynthTab~\cite{Zang2024synthtab} and is not a comparison against it.
It is worth noting, however, that we used considerably less synthetic data than SynthTab (we only synthesized 360 tracks to train each model, while SynthTab added 6,700 hours \cite{Zang2024synthtab}).
In future work, we will systematically explore the impacts of the various factors that go into simulating guitar performances, such as including real versus computer-generated tones and/or effects, the impact of delay effects that repeat the onset of a tone, and the amount of data used for training.

\section{Conclusion}


We have demonstrated the impact that using synthetic guitar performances as training data has on the robustness of GTT models.
Specifically, we leveraged guitar tablatures to produce these performances using real recordings of electric guitar notes with a wide array of processing that included real audio effects hardware.
We showed increased model robustness on multi-pitch and tablature prediction metrics via our proposed method. 
In the future, we look forward to enhancing datasets for GTT using our methodology and systematically studying all the parameters involved in the data generation, such as dataset size or tone diversity. 

\section{Acknowledgements}
This work was supported by the IEEE Mentoring Experiences for Underrepresented Young Researchers (ME-UYR) and Mexico's National Scholarship for Graduate Studies from the National Council of Humanities, Sciences, and Technologies (CONAHCYT). The authors thank the funding sources and collaborators, particularly PhD candidate Elliot Hernandez, who proofread this manuscript.

\bibliographystyle{IEEEtran}
\bibliography{DAFx24_tmpl} 

\begin{thebibliography}{10}
\providecommand{\url}[1]{#1}
\csname url@samestyle\endcsname
\providecommand{\newblock}{\relax}
\providecommand{\bibinfo}[2]{#2}
\providecommand{\BIBentrySTDinterwordspacing}{\spaceskip=0pt\relax}
\providecommand{\BIBentryALTinterwordstretchfactor}{4}
\providecommand{\BIBentryALTinterwordspacing}{\spaceskip=\fontdimen2\font plus
\BIBentryALTinterwordstretchfactor\fontdimen3\font minus \fontdimen4\font\relax}
\providecommand{\BIBforeignlanguage}[2]{{%
\expandafter\ifx\csname l@#1\endcsname\relax
\typeout{** WARNING: IEEEtran.bst: No hyphenation pattern has been}%
\typeout{** loaded for the language `#1'. Using the pattern for}%
\typeout{** the default language instead.}%
\else
\language=\csname l@#1\endcsname
\fi
#2}}
\providecommand{\BIBdecl}{\relax}
\BIBdecl

\bibitem{benetos2018automatic}
E.~Benetos, S.~Dixon, Z.~Duan, and S.~Ewert, ``Automatic music transcription: An overview,'' \emph{IEEE Signal Processing Magazine}, vol.~36, no.~1, pp. 20--30, 2018.

\bibitem{Gowrishankar2016amtreview}
B.~S. Gowrishankar and N.~U. Bhajantri, ``An exhaustive review of automatic music transcription techniques: Survey of music transcription techniques,'' in \emph{International Conference on Signal Processing, Communication, Power and Embedded System}, 2016, pp. 140--152.

\bibitem{burlet2013robotaba}
G.~Burlet and I.~Fujinaga, ``Robotaba guitar tablature transcription framework.'' in \emph{ISMIR}, 2013, pp. 517--522.

\bibitem{Peleaari2008Multimodal}
M.~Paleari, B.~Huet, A.~Schutz, and D.~Slock, ``A multimodal approach to music transcription,'' 11 2008, pp. 93 -- 96.

\bibitem{thompson2011speaking}
D.~E. Thompson, ``Speaking their language: Guitar tablature in the middle school classroom,'' \emph{General Music Today}, vol.~24, no.~3, pp. 53--57, 2011.

\bibitem{harrison2010challenges}
E.~Harrison, ``Challenges facing guitar education,'' \emph{Music educators journal}, vol.~97, no.~1, pp. 50--55, 2010.

\bibitem{gavito2015oral}
C.~M. Gavito, ``Oral transmission and the production of guitar tablature books in seventeenth-century italy,'' \emph{Recercare}, pp. 185--208, 2015.

\bibitem{de2022guitar}
J.~De~Souza, ``Guitar thinking,'' \emph{Soundboard Scholar}, vol.~7, no.~1, p.~1, 2022.

\bibitem{de2018fretboard}
------, ``Fretboard transformations,'' \emph{Journal of Music Theory}, vol.~62, no.~1, pp. 1--39, 2018.

\bibitem{koozin2011pop}
T.~Koozin, ``Guitar voicing in pop-rock music: A performance-based analytical approach,'' \emph{Music Theory Online}, vol.~17, 10 2011.

\bibitem{xi2018guitarset}
Q.~Xi, R.~Bittner, J.~Pauwels, X.~Ye, and J.~Bello, ``Guitarset: A dataset for guitar transcription.'' in \emph{ISMIR}, 2018.

\bibitem{wiggins2019guitar}
A.~Wiggins and Y.~Kim, ``Guitar tablature estimation with a convolutional neural network.'' in \emph{ISMIR}, 2019, pp. 284--291.

\bibitem{Cwitkowitz2023Fretnet}
F.~Cwitkowitz, T.~Hirvonen, and A.~Klapuri, ``Fretnet: Continuous-valed pitch contour streaming,'' in \emph{IEEE International Conference on Acoustics, Speech and Signal Processing}, 2023, pp. 1--5.

\bibitem{kim2022attentionMechanism}
S.~Kim, T.~Hayashi, and T.~Toda, ``Note-level automatic guitar transcription using attention mechanism,'' in \emph{European Signal Processing Conference}, 2022, pp. 229--233.

\bibitem{bastas2022inharmonicity}
G.~Bastas, S.~Koutoupis, M.~Kaliakatsos-Papakostas, V.~Katsouros, and P.~Maragos, ``A few-sample strategy for guitar tablature transcription on inharmonicity analysis \& playability constraints,'' in \emph{IEEE International Conference on Acoustics, Speech and Signal Processing}, 2022, pp. 771--775.

\bibitem{byambatsogt2020robots}
G.~Byambatsogt, L.~Choimaa, and G.~Koutaki, ``Guitar chord sensing and recognition using multi-task learning and physical data augmentation with robotics,'' \emph{Sensors}, vol.~20, no.~21, 2020.

\bibitem{riley2024high}
X.~Riley, D.~Edwards, and S.~Dixon, ``High resolution guitar transcription via domain adaptation,'' 2024.

\bibitem{Jadhav2022}
Y.~Jadhav, A.~Patel, R.~H. Jhaveri, and R.~Raut, ``Transfer learning for audio waveform to guitar chord spectrograms using the convolution neural network,'' \emph{Mobile Information Systems}, vol. 2022, p. 8544765, Aug 2022.

\bibitem{Zang2024synthtab}
Y.~Zang, Y.~Zhong, F.~Cwitkowitz, and Z.~Duan, ``Synthtab: Leveraging synthesized data for guitar tablature transcription,'' 04 2024, pp. 1286--1290.

\bibitem{quinonero2022dataset}
J.~Qui{\~n}onero-Candela, M.~Sugiyama, A.~Schwaighofer, and N.~D. Lawrence, \emph{Dataset shift in machine learning}.\hskip 1em plus 0.5em minus 0.4em\relax Mit Press, 2022.

\bibitem{shorten2019survey}
C.~Shorten and T.~M. Khoshgoftaar, ``A survey on image data augmentation for deep learning,'' \emph{Journal of big data}, vol.~6, no.~1, pp. 1--48, 2019.

\bibitem{salamon2017scaper}
J.~Salamon, D.~MacConnell, M.~Cartwright, P.~Li, and J.~P. Bello, ``Scaper: A library for soundscape synthesis and augmentation,'' in \emph{Workshop on Applications of Signal Processing to Audio and Acoustics}.\hskip 1em plus 0.5em minus 0.4em\relax IEEE, 2017, pp. 344--348.

\bibitem{roman2024spatial}
I.~R. Roman, C.~Ick, S.~Ding, A.~S. Roman, B.~McFee, and J.~P. Bello, ``Spatial scaper: a library to simulate and augment soundscapes for sound event localization and detection in realistic rooms,'' in \emph{IEEE International Conference on Acoustics, Speech and Signal Processing}.\hskip 1em plus 0.5em minus 0.4em\relax IEEE, 2024.

\bibitem{roman2024enhanced}
A.~S. Roman, B.~Balamurugan, and R.~Pothuganti, ``Enhanced sound event localization and detection in real 360-degree audio-visual soundscapes,'' \emph{arXiv preprint arXiv:2401.17129}, 2024.

\bibitem{dinkel2021towards}
H.~Dinkel, M.~Wu, and K.~Yu, ``Towards duration robust weakly supervised sound event detection,'' \emph{IEEE/ACM Transactions on Audio, Speech, and Language Processing}, vol.~29, pp. 887--900, 2021.

\bibitem{pedroza2022egfxset}
H.~Pedroza, G.~Meza, and I.~Roman, ``Egfxset: Electric guitar tones processed through real effects of distortion, modulation, delay and reverb,'' \emph{ISMIR LBD}, 2022.

\bibitem{sarmento2021dadagp}
P.~Sarmento, A.~Kumar, C.~Carr, Z.~Zukowski, M.~Barthet, and Y.-H. Yang, ``Dadagp: a dataset of tokenized guitarpro songs for sequence models,'' in \emph{ISMIR}, 2021.

\end{thebibliography}

\end{document}